\documentclass[preprint]{aastex631}
\shorttitle{Which Component of Solar Magnetic Field Drives IMF Evolution?}
\shortauthors{Yoshida et al.}
\graphicspath{{./}{figures/}}

\begin{document}

\title{Which Component of Solar Magnetic Field Drives the Evolution of Interplanetary Magnetic Field over Solar Cycle?}

\author{Minami Yoshida}
\affiliation{Department of Earth and Planetary Science, The University of Tokyo, 7-3-1, Hongo, Bunkyo-ku, Tokyo 113-0033, Japan}
\affiliation{Institute of Space and Astronautical Science, Japan Aerospace Exploration Agency, 3-1-1, Yoshinodai, Chuo-ku, Sagamihara, Kanagawa 252-5210, Japan}

\author{Toshifumi Shimizu}
\affiliation{Institute of Space and Astronautical Science, Japan Aerospace Exploration Agency, 3-1-1, Yoshinodai, Chuo-ku, Sagamihara, Kanagawa 252-5210, Japan}
\affiliation{Department of Earth and Planetary Science, The University of Tokyo, 7-3-1, Hongo, Bunkyo-ku, Tokyo 113-0033, Japan}

\author{Shin Toriumi}
\affiliation{Institute of Space and Astronautical Science, Japan Aerospace Exploration Agency, 3-1-1, Yoshinodai, Chuo-ku, Sagamihara, Kanagawa 252-5210, Japan}



\begin{abstract}
The solar magnetic structure changes over the solar cycle. It has a dipole structure during solar minimum, where the open flux extends mainly from the polar regions into the interplanetary space. During maximum, a complex structure is formed with low-latitude active regions and weakened polar fields, resulting in spread open field regions. However, the components of the solar magnetic field that is responsible for long-term variations in the interplanetary magnetic field (IMF) are not clear,
and the IMF strength estimated based on the solar magnetic field is known to be underestimated by a factor of 3 to 4 against the actual in-situ observations (the open flux problem). To this end, we decomposed the coronal magnetic field into the components of the spherical harmonic function of degree and order $(\ell, m)$ using the potential field source surface model with synoptic maps from SDO/HMI for 2010 to 2021. As a result, we found that the IMF rapidly increased in December 2014 (seven months after the solar maximum), which coincided with the increase in the equatorial dipole, $(\ell, m)=(1, \pm1)$, corresponding to the diffusion of active regions toward the poles and in the longitudinal direction. The IMF gradually decreased until December 2019 (solar minimum) and its variation corresponded to that of the non-dipole component $\ell\geq2$.
Our results suggest that the understanding of the open flux problem may be improved by focusing on the equatorial dipole and the non-dipole component and that the influence of the polar magnetic field is less significant.
\end{abstract}

\keywords{Sun:heliosphere --- Sun:corona --- Sun:photosphere --- Sun:magnetic field }


\section{Introduction} \label{sec:intro}

The global structure of the solar magnetic field changes over the solar cycle.
When magnetic activity is low during the solar minimum, a unipolar magnetic field extends from the polar regions where coronal holes are widely distributed \citep{1973SoPh...29..505K} forming a dipole structure.
In the solar maximum, the number of active regions increases, which makes the magnetic field structure complex. Changes in coronal structure are well captured observationally by coronagraphs  (\citealp{1992SSRv...61..393K}; \citealp{2019ApJ...883..152D}) and during the solar eclipses (\citealp{2015ApJ...800...90P}; \citealp{2018NatAs...2..913M}). They are also predicted using coronal field modeling \citep{2014SoPh..289..631Y}.

Some coronal magnetic fields extend to several solar radii and further into the interplanetary space called ``open" fields. The fast solar wind is believed to flow out through the open field toward the interplanetary space.
The open field extends mainly from the coronal holes, according to the standard coronal structure \citep{2012LRSP....9....6M}.
In the outer corona far from the photosphere, the solar wind plasma with the solar magnetic field is pushed out into the heliosphere \citep{1958ApJ...128..664P}. In particular, fast solar winds flow out from the polar coronal holes into the interplanetary space along the open magnetic field \citep{1998GeoRL..25....1M}.
The connection of the solar magnetic field to the interplanetary magnetic field (IMF) is essential for understanding the magnetic structure in the interplanetary space, which changes over the solar activity cycle. The strength of the IMF and solar wind velocity have been observed in situ near the Earth since 1965 \citep{2013LRSP...10....5O}. Currently, the Wind \citep{1995SSRv...71..207L} and Advanced Composition Explorer (ACE; \citealp{2005GeoRL..3215S01B}) satellites measure them at the Lagrangian point 1 (L1), which are publicly available in the OMNI dataset \citep{https://doi.org/10.1029/2004JA010649}.
The IMF strength can be estimated from measurements of the magnetic field in the solar photosphere by extrapolating the field lines with some assumptions.
The estimated magnetic field should agree with the strength of the IMF measured near the Earth. However, \citet{2017ApJ...848...70L} found that the extrapolated field strength is two to four times smaller than that obtained by the in-situ measurements, regardless of the photospheric magnetic field data and the model used in the estimation. This quantitative discrepancy is known as ``the open flux problem" \citep{2017ApJ...848...70L}.

Several possible causes of this discrepancy have been considered and discussed: 1) the open magnetic flux is believed to be mainly rooted into the coronal holes but regions outside coronal holes may also contribute to the open flux; 2) the polar photospheric field strength is underestimated because it is difficult to measure the magnetic flux density with high accuracy in the polar regions; and 3) we have not found a good assumption for the coronal magnetic field modeling.
For 1), the open magnetic fields are believed to be rooted primarily in coronal holes, but \citet{2019SoPh..294...19W} pointed out that this may not be the case. They manually identified coronal holes using the extreme ultraviolet (EUV) images and evaluated the amount of open field observationally. However, even with this method, the estimated open magnetic field was underestimated compared to the observed values at 1 AU. This indicates that source regions of open flux may exist in addition to coronal holes, which appear dark in EUV. Some studies have attempted to identify coronal holes more clearly; \citet{2019SoPh..294..144H} proposed a new coronal hole identification method using an intensity-based threshold method and estimated coronal hole boundaries. \citet{2021ApJ...918...21L} calculated open flux using several coronal hole identifications including the new method above, but the discrepancy remains.

For 2), \citet{2019ApJ...884...18R} suggested that the open flux problem may be improved when the polar magnetic field is observed more accurately. Assuming that the polar magnetic field, which is difficult to observe from our vantage point in the ecliptic plane, is underestimated, they made new magnetographs with additional magnetic flux in polar regions and found that the estimated amount of the open magnetic field with the new magnetograph almost matches that of the in-situ measurement. However, the amount of flux added to the new map seems large compared to results from Hinode's observations, the highest resolution observations for the polar regions so far (\citealp{2008ApJ...688.1374T};  \citealp{2010ApJ...719..131I}). This study was carried out only for one solar rotation near the solar minimum. Therefore, the results in the other phases are not clear.
\citet{2022ApJ...926..113W} introduced a new Zeeman saturation correction to improve the magnetograms by Mount Wilson Observatory and Wilcox Solar Observatory, and obtained the open flux sufficient to explain the underestimation.
However, it is uncertain whether a similar correction is suitable for space-borne magnetograms.

Furthermore, \citet{2021A&A...650A..18B} and \citet{2021A&A...650A..19R} suggested that the open flux problem may occur not only in the near-Earth regions but also in regions closer to the Sun. They analyzed the results of the FIELDS instrument on the Parker Solar Probe, which performed in-situ measurements from 1 AU to 0.13 AU and compared them with the magnetic field estimated using the coronal magnetic models. A discrepancy was found between the estimated and observed magnetic fields, even at a distance of 0.13 AU from the Sun.
The results suggest that we should investigate the origins of the problem in the areas close to the Sun. \citet{2021A&A...650A..19R} argued that the assumptions and boundary conditions of the coronal field models should also be further examined.

As described above, the open magnetic field estimated from the photospheric magnetic field is underestimated compared to the in-situ measurements of IMF. However, the temporal variation in this underestimation over the entire solar cycle has not been well studied.
It is important to study the open flux problem over the solar cycle because the coronal field structure changes as a function of phase in the cycle.
Previous studies as listed above have suggested the possibility that the source of the open flux problem lies in
regions close to the Sun. Therefore, our study compares the open flux extending from the photospheric magnetic field with the strength of the IMF near Earth over the entire solar cycle using data from 2010 to 2021. According to \citet{2022ApJ...926..113W}, the main contribution to the open flux comes from the dipole or quadrupole magnetic field.
Therefore, we first compare the variability of the total unsigned solar photospheric flux with the near-Earth open flux and then investigate the variability of the dipole and non-dipole components of the photospheric field to understand which component of the magnetic field in the photosphere is responsible for the open flux as it changes with the solar cycle.

In Section \ref{sec:Analysis}, we describe the observations and data analysis. In Section \ref{sec:results}, we present the results of our analysis, followed by discussion in Section \ref{sec:Discussion} and conclusions in Section \ref{sec:Conclusion}.

\section{Observations and data analysis}\label{sec:Analysis}

In this study, we compared the open flux using a coronal field model with IMF by in-situ measurement from May 2010 to October 2021. The data used in the comparisons and the methods of data analysis are described below.

\subsection{In-situ measurement of the IMF}
We used the in-situ measurement from Advanced Composition Explorer (ACE; \citealp{2005GeoRL..3215S01B}) and Wind \citep{1995SSRv...71..207L}, which are available as Low-Resolution OMNI (\url{https://omniweb.gsfc.nasa.gov/form/dx1.html}).
The data have already been extensively cross-compared and cross-normalized for some spacecraft and parameters. High-resolution data from satellites in the L1 orbit are time-shifted to the expected magnetospheric arrival time before taking the 1-hour average (Low-Resolution OMNI data).
For comparison with the solar magnetic field (Section \ref{sec:ob-open_flux}), we took the absolute values of the radial component and averaged them over each Carrington rotation. Each Carrington rotation is defined by one solar rotation period labeled as Carrington rotation number (CR) since 1853.
In this study, we examined the data in the period from CR2097 (May 2010) to CR2249 (October 2021).

\subsection{Solar photospheric magnetogram and  the open flux estimate}\label{sec:ob-open_flux}
To estimate the value of the open flux at 1 AU far from the Sun $|B_{r 1 \rm{AU}}|$, we used the potential field source surface (PFSS) model (\citealp{1969SoPh....9..131A}; \citealp{1969SoPh....6..442S}) and the photospheric magnetic map as a boundary condition.
The PFSS model is one of the simplest coronal magnetic models, which has two major assumptions.
First, it assumes a potential field with no electric current and neglects the Lorentz force.
The second assumption is that all the magnetic fields are oriented in the radial direction at the source surface, which is the upper boundary.
The PFSS code for modeling the three-dimensional (3D) coronal field used in this study was developed by M. L. DeRosa (\url{https://www.lmsal.com/~derosa/pfsspack}).

We used synoptic magnetograms from the Helioseismic and Magnetic Imager (HMI) onboard the Solar Dynamics Observatory (SDO) spacecraft (\citealp{2012SoPh..275....3P}; \citealp{2012SoPh..275..207S}).
The spatial distribution of magnetic flux density on the solar surface has been observed by SDO/HMI since May 2010. The HMI observes the full solar disk in the Fe I absorption line at 6173 {\AA} with a spatial sampling of 0.5\arcsec.
Several types of data are available, and we used synoptic maps created from line-of-sight magnetograms available at the Joint Science Operations Center (JSOC; \dataset[http://jsoc.stanford.edu]{http://jsoc.stanford.edu})
and simply filled the polar regions with no observation with 0 G.
Synoptic maps are 2D maps with Carrington longitude on the horizontal axis covering the entire solar surface, which allows for the extrapolation of the 3D coronal magnetic field by the PFSS model. Each map has 3600 pixels in the horizontal direction and 1440 pixels with a sine latitude in the vertical direction.

We used the radial magnetic component $B_{r}$ available in the synoptic maps to model the coronal magnetic field and reduced the size of the maps to $600\times300$ pixels each to accelerate the calculation speed. When the grid size changed from coarse to fine, the open flux at the source surface $\Phi_{\rm{open}}$ fluctuated. We checked $\Phi_{\rm{open}}$ by changing the map size, and confirmed that a reduced number of $600\times300$ pixels provided stable results with fewer fluctuations.

The calculations of $\Phi_{\rm{open}}$ and $|B_{r 1 \rm{AU}}|$ are as follows: First, based on the extrapolated coronal magnetic field, we calculated the unsigned magnetic flux $\Phi_{\rm{open}}$ in the radial direction at the source surface. We set 2.5 solar radii as the source surface for the upper boundary.

When using the coronal magnetic field model, 2.5 solar radii have been empirically set as the source surface.
According to \citet{2019JGRA..124.8280A}, when the source surface set to 2.3 to 2.6 solar radii over the solar cycle, the regions of coronal holes and open fields agree only about 60\%. On the other hand, when the height of the source surface is small, the agreement of their regions increases, but open field regions exist outside of the coronal holes. This indicates that we cannot simply determine the source surface.
\citet{2017ApJ...848...70L} changed the source surface from 1.3 to 2.5 solar radii and compared the open flux area with the area of the coronal holes. As a result, the area of the open field calculated by 2.5 solar radii or 2.0 solar radii are in good agreement with that of coronal holes. Following this result, we use 2.5 solar radii in this study.

Considering the spherical coordinate system $(r, \theta, \phi)$ with the center of the Sun as the origin, we obtained the open flux at the source surface as
\begin{eqnarray}
\Phi_{\rm{open}}=\int_{R_{\rm{ss}}}\left|B_r\right|dS=\sum_{ij} |B_{r} (i, j)| \Delta S, \\
\Delta S=R_{\rm{ss}}^2\sin{\theta}\Delta\theta\Delta\phi,
\end{eqnarray}
where $R_{\rm{ss}}$ is the distance from the center of the Sun to the source surface, and $i$ and $j$ represent the grid position in the reduced maps.
The open flux at 1 AU away from the Sun, $|B_{r 1 \rm{AU}}|$, was estimated using the following equation, assuming that the value of IMF is independent of heliospheric latitude (\citealp{1995ApJ...447L.143W}; \citealp{2008JGRA..11312103O}):
\begin{eqnarray}
|B_{r 1\rm{AU}}| &=& \frac{|\Phi_{\rm{open}}|}{4\pi r_{1\rm{AU}}^2} \nonumber\\
&=& \frac{1}{4\pi}(\frac{R_{\rm{ss}}}{215})^2\sum_{ij} |B_{r} (i, j)|\sin{\theta}\Delta\theta\Delta\phi,
\end{eqnarray}
where $r_{1\rm{AU}}$ is the distance between the Sun and Earth ($215$ solar radii) .

Spherical harmonic expansion was applied to each synoptic map of the photosphere to derive the coronal magnetic field (see Appendix \ref{appendix} for the equations). The degree $\ell$ and order $m$ characterize the spherical harmonic functions and, in this study, we varied $\ell$ to separate the modes of the solar magnetic field. For example, we set $\ell_{\rm{max}}=300$ when we calculated the coronal magnetic field including fine structures, whereas we calculated up to $\ell=1$ when estimating the dipole magnetic field.

\section{Results}\label{sec:results}

\subsection{Temporal evolutions of measured and extrapolated open fluxes}\label{sec:re_photo}
\begin{figure}[ht!]
\begin{center}
\includegraphics[width=0.7\textwidth]{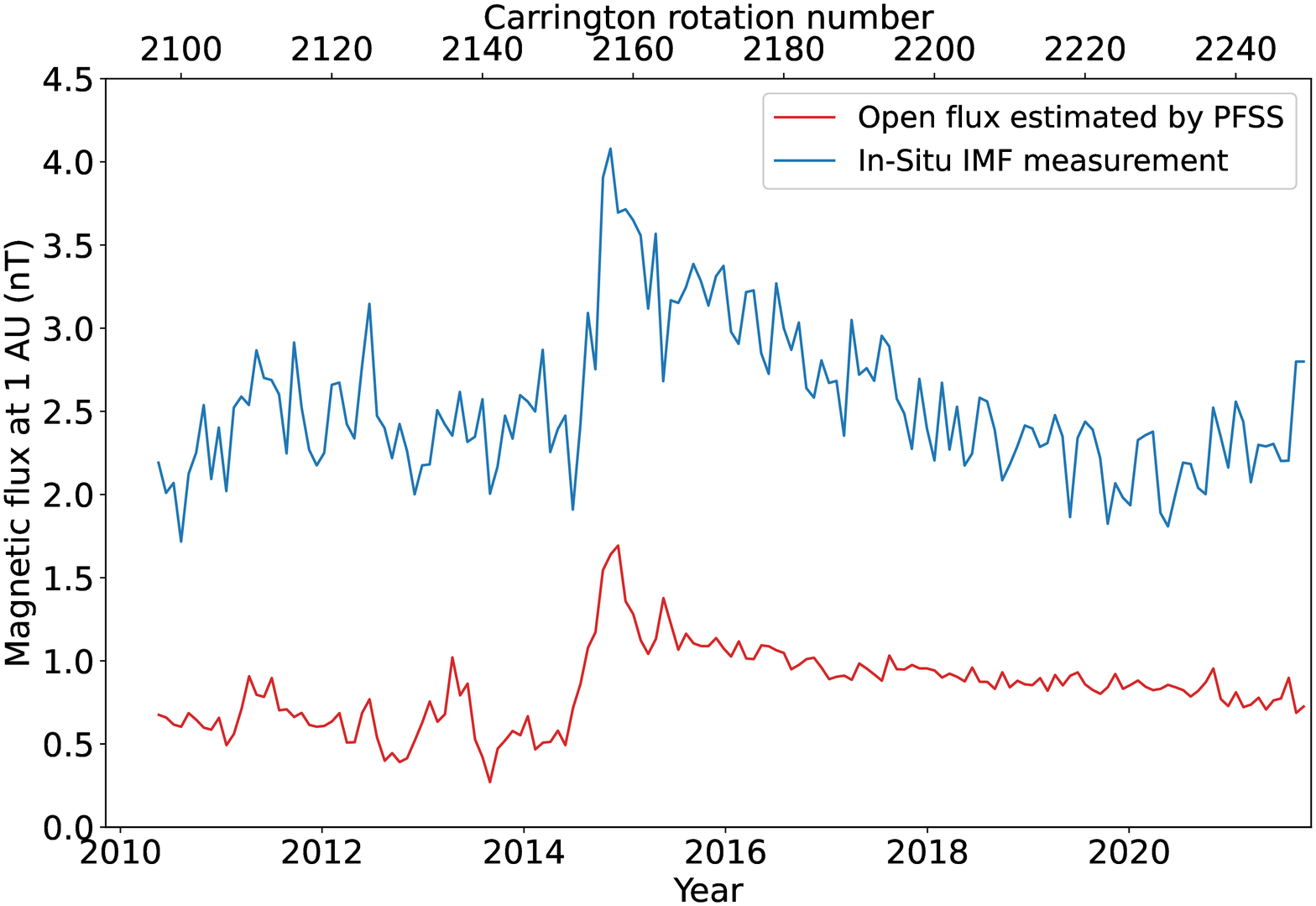}
  \includegraphics[width=0.7\textwidth]{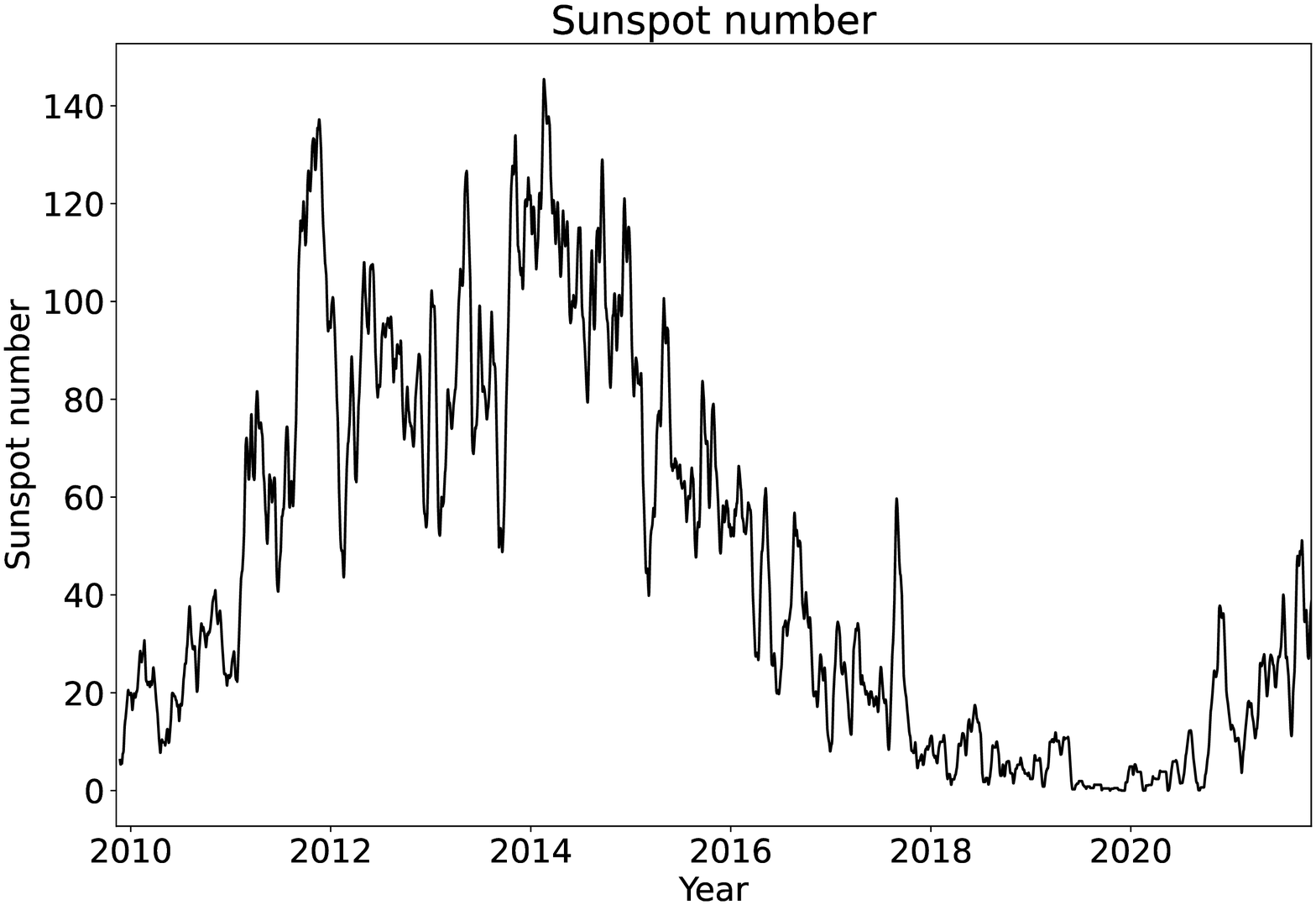}
 \end{center}
  \caption{The top panel shows the temporal variations of IMF (blue line) and the open flux extrapolated from the solar photosphere $|B_{r 1\rm{AU}}|$ (red line) during 2010 to 2021. Each data point corresponds to the data for the Carrington rotation. The bottom is the time variation of the monthly mean sunspot number.}
  \label{pic:problem}
 \end{figure}

We compared the extrapolated open flux $|B_{r 1\rm{AU}}|$ to the IMF by in-situ measurements near Earth. Figure \ref{pic:problem} shows $|B_{r 1\rm{AU}}|$ and the IMF as a function of time in 2010-2021. The temporal evolution of the monthly mean sunspot number is also provided to show the timing of the solar maximum.
The solar maximum is in April 2014 and the minima are in December 2008 and December 2019 (\url{https://www.sidc.be/silso/cyclesminmax}).
We can confirm that the extrapolated open flux $|B_{r 1\rm{AU}}|$ is underestimated by a factor of three from the IMF value, and the underestimation is always seen independent of the phase of the solar cycle. The in-situ measurement of the IMF at 1 AU showed a constant flux of around 2.4 nT with fluctuations from 2010 to 2014. The IMF suddenly increased from 2.4 to 4.1 nT in November 2014 (CR2157) and it gradually decreased until 2020, at the solar minimum.
The timing of the peaked IMF lagged seven months after the peak of the sunspot number in April 2014. The open flux shows a similar temporal behavior; it peaked in a short time scale in November 2014 and then showed a gradual decrease.
Coronal mass ejections (CMEs) usually increase the magnetic field near the Earth and are measured more abundantly during the maximum period \citep{2012LRSP....9....3W}. However, because CMEs are transient phenomena, these effects are smoothed out by averaging over each Carrington rotation. Therefore, it is unlikely that the observed enhancement in the IMF is owing to the CMEs.

\begin{figure}[ht!]
\begin{center}
\includegraphics[width=0.7\textwidth]{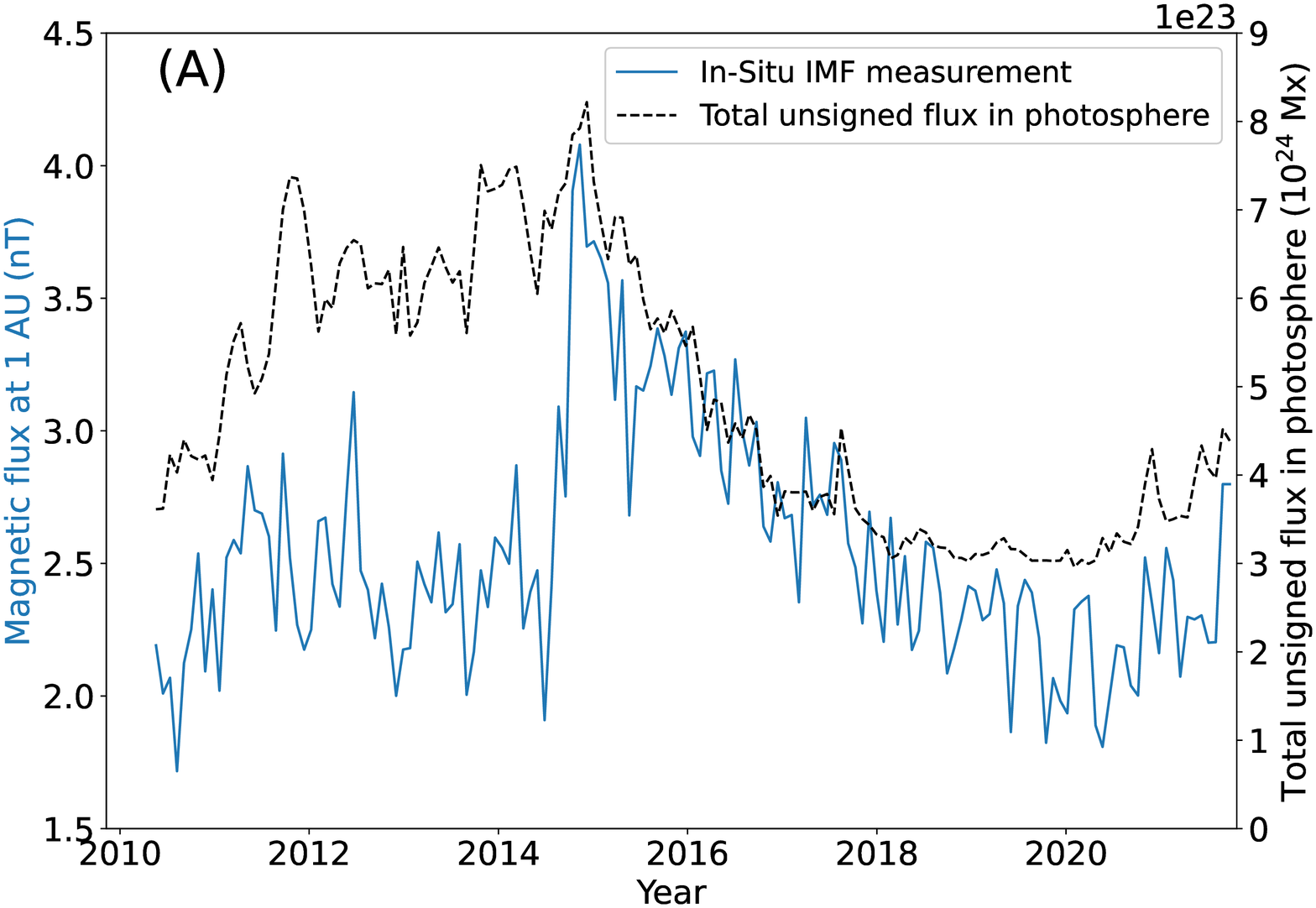}
  \includegraphics[width=0.7\textwidth]{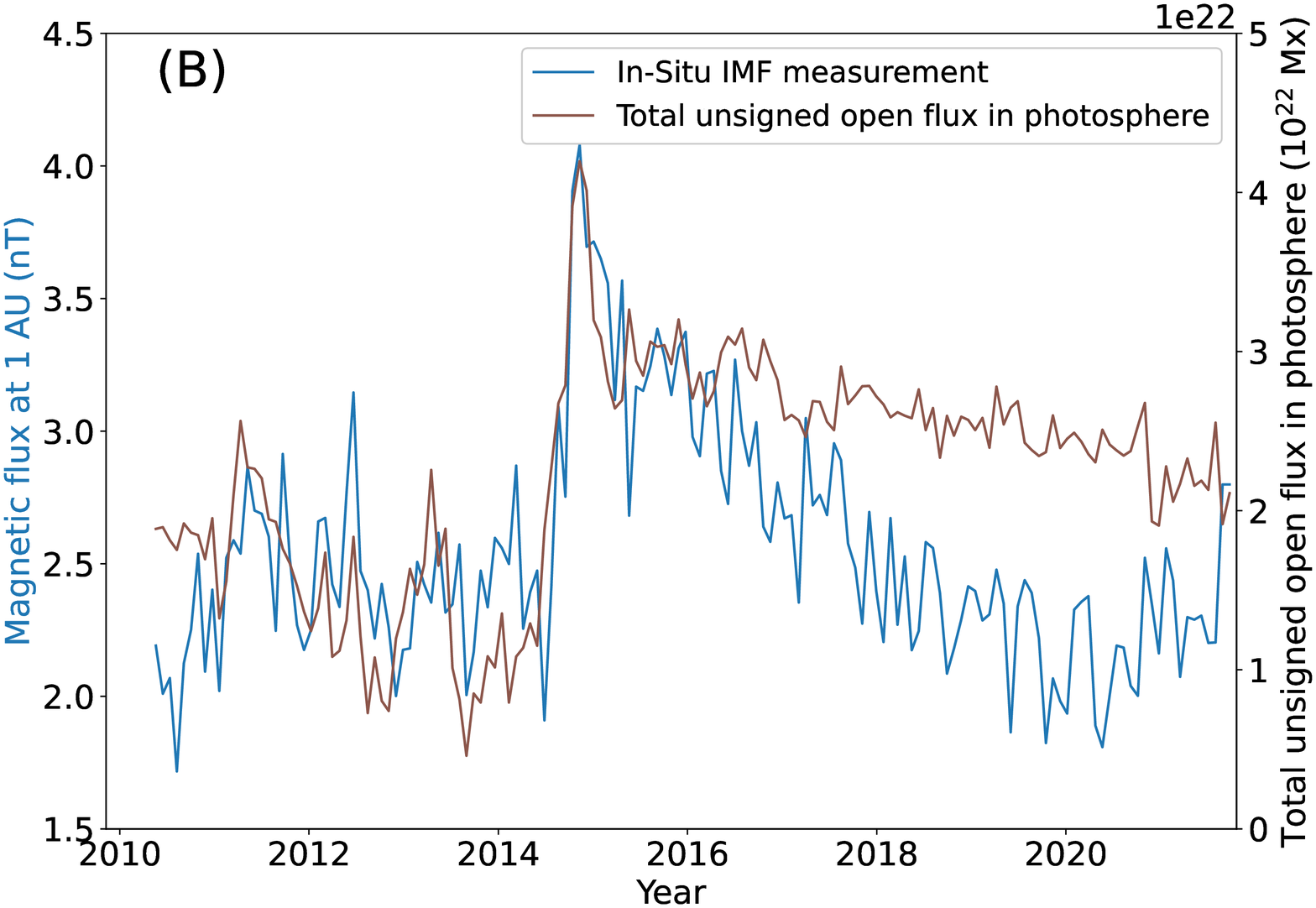}
 \end{center}
  \caption{(A) The temporal variation of the total unsigned magnetic flux at the solar photosphere (black dashed line, the right axis) compared with that of IMF (blue line, the left axis) over the solar cycle (2010 to 2021). (B) The temporal variations of the total unsigned open flux at the photosphere only containing the open flux (brown line, the right axis) during 2010 to 2021, compared with that of the IMF. Each data point corresponds to the data for the Carrington rotation. }
  \label{pic:open}
 \end{figure}

Figure \ref{pic:open} (A) shows the temporal variations in the total unsigned magnetic flux in the photosphere over the solar cycle, which is compared to the radial in-situ flux at 1 AU. The result shows that the variation in the photospheric total flux is poorly correlated with the radial flux at 1 AU before 2015 (the Pearson correlation coefficient is 0.47; Table \ref{table:pcc}), whereas they show a stronger correlation with each other after 2015 (the correlation coefficient is 0.87). The correlation between the two is 0.45 for the entire period. The transition from poor to strong correlation occurs at the peak of the open flux in CR2157 (from November to December 2014).
Figure \ref{pic:open} (B) shows temporal variations of the total unsigned open flux ($|\Phi_\mathrm{open}|$) in the photosphere.
The open field area in the photosphere is defined as the footpoints of the magnetic field lines that reach the source surface by the PFSS extrapolation. This is obtained by tracing the magnetic field lines backwards from the source surface to the photosphere.
We can see that the peak of in-situ measurement around 2015 coincides with the peak of the open field in the photosphere. In addition, the trends of both fluctuations are in good agreement between 2010 and 2015, i.e., before the IMF peak. The correlation coefficient is 0.61 for this period. Conversely, these trends are less consistent after 2015. The decreasing trend of the observed IMF deviates from that of the total open flux in the photosphere. The correlation coefficient goes down to 0.20 in the period from January 2017 to October 2021. From CR2157 to CR2231 (November 2014 to June 2020), in the period from the IMF maximum to the minimum, IMF is down to $44\%$ of the peak, while the total open unsigned flux only reaches $60\%$.
Thus, the amount of the open field in the photosphere is not perfectly proportional to the open flux near Earth.

\begin{table}[!ht]
     \centering
     \caption{Pearson correlation coefficients between various magnetic fluxes and IMF.}
     \begin{tabular}{c|cccc}
     \hline
         Carrington rotation & CR2097-CR2249 & CR2097-CR2157 & CR2157-CR2249 & CR2186-CR2249 \\
         Year & 2010-2021 & 2010-2015 & 2015-2021 & 2017-2021 \\ \hline
         Total flux in the photosphere & 0.45 & 0.47 & 0.87 & 0.45 \\
         Total open flux & 0.50 & 0.61 & 0.65 & 0.20 \\
         Total dipole flux & 0.68 & 0.63 & 0.74 & 0.29 \\
         Total non-dipole flux & 0.42 & 0.43 & 0.87 & 0.44 \\
         Flux ($-45^{\circ}$ to $45^{\circ}$) & 0.23 & 0.06 & 0.85 & 0.61 \\
     \hline
     \end{tabular}
     \label{table:pcc}
 \end{table}

\subsection{Open flux and the dipole structure of the Sun}\label{sec:re_dipole}

\begin{figure}[ht!]
  \begin{center}
  \includegraphics[width=0.7\textwidth]{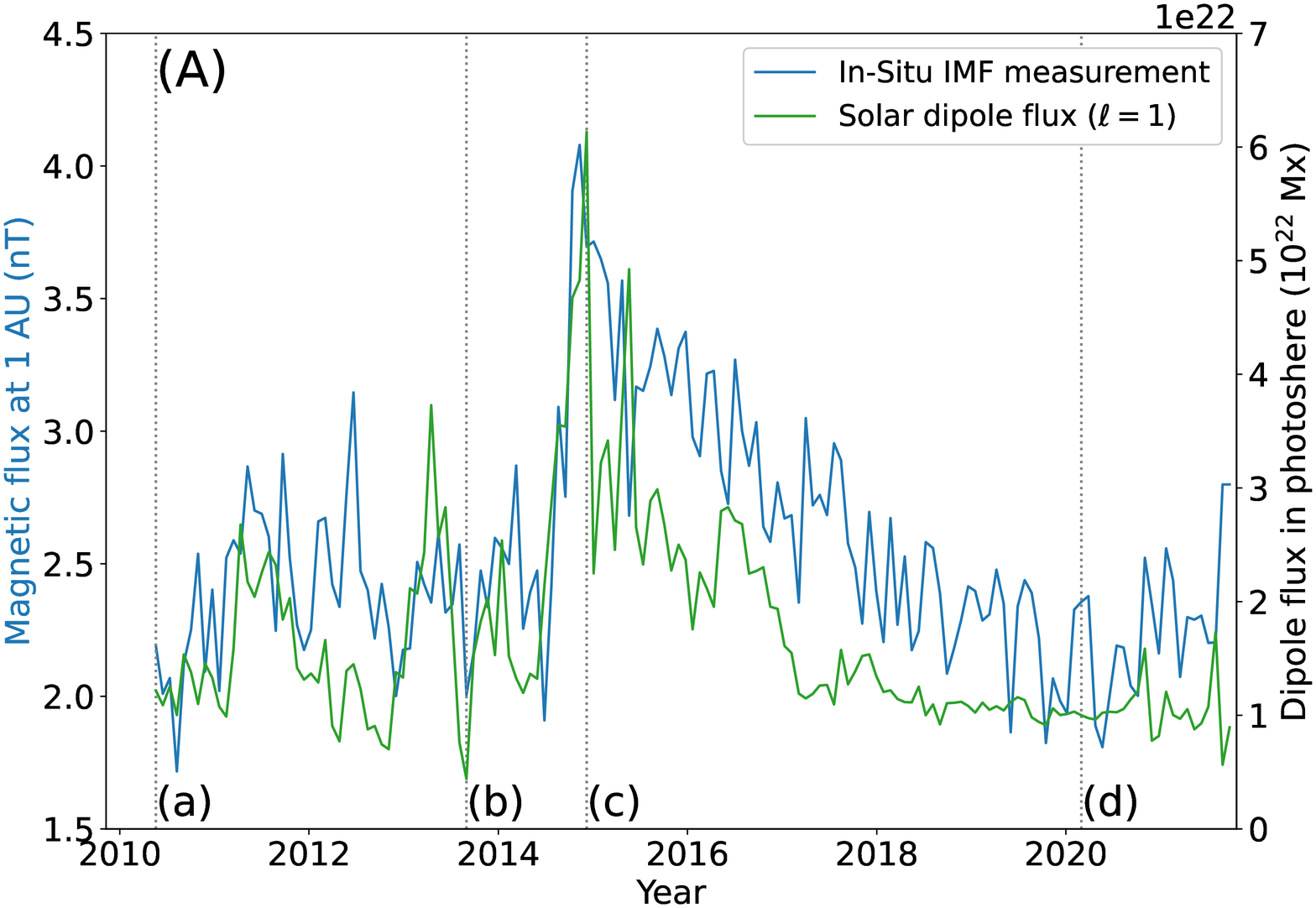}
  \includegraphics[width=0.7\textwidth]{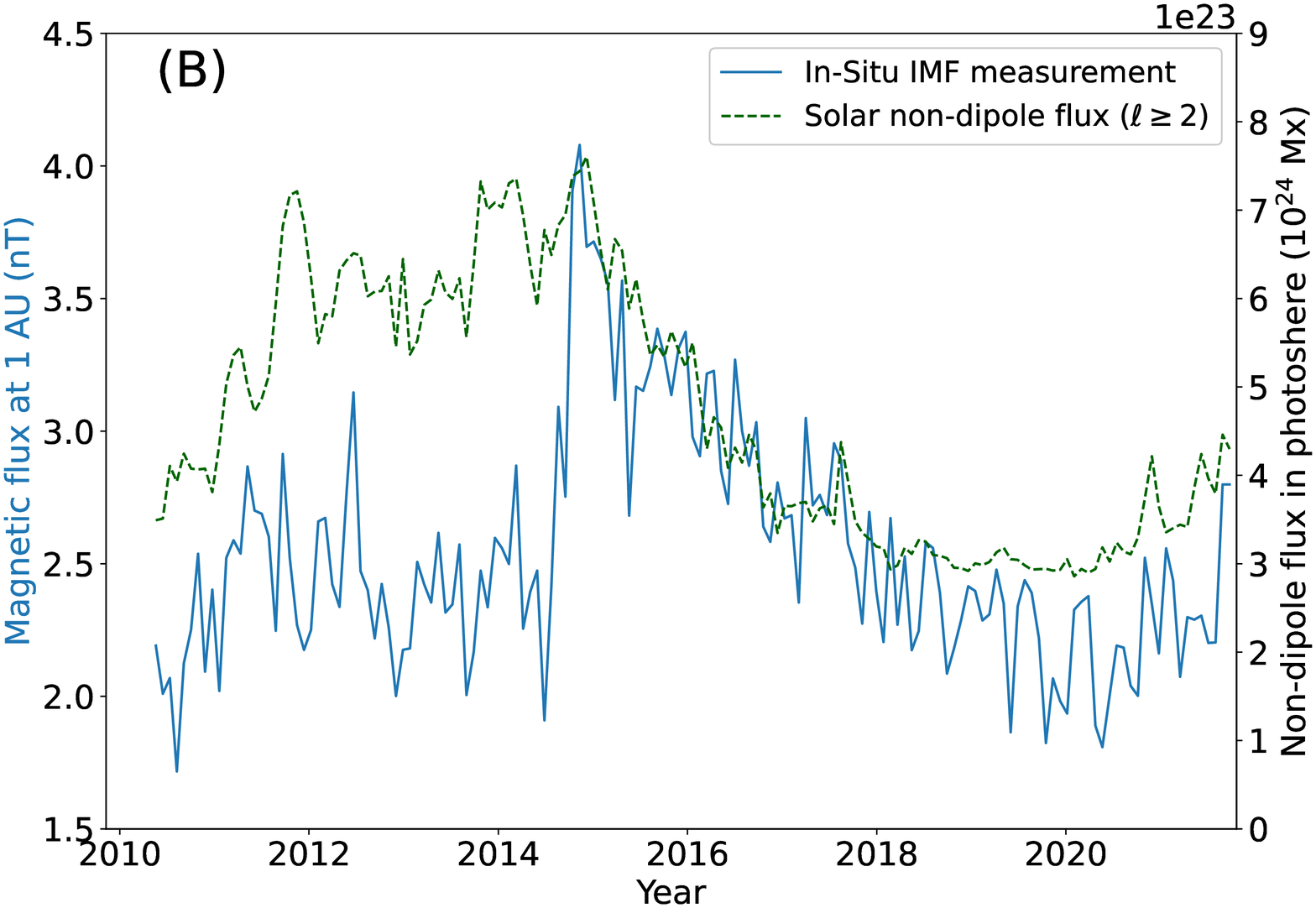}
 \end{center}
  \caption{(A) The temporal variations of total dipole flux (green line, the right axis) in 2010 to 2021 compared with that of IMF (blue line, the left axis). The timing of the peak coincides with each other. The dotted lines indicate the time of the maps shown in Figure \ref{pic:2d_l1}. (B) The temporal variations of solar total non-dipole flux (green dashed line, the right axis) during 2010 to 2021, compared to those of the IMF (blue line, the left axis). The timing of both peaks coincides. Each data point corresponds to the data for the Carrington rotation. }
  \label{pic:di_nondi}
 \end{figure}

In Figure \ref{pic:di_nondi} (A), the green line shows the variation in the total solar dipole flux from 2010 to 2021 compared to the time variations of the IMF at 1 AU.
 The solar total dipole flux was derived by calculating Equation (\ref{eq:SHF1}) with $\ell=1$.
The solar total dipole flux increased rapidly in December 2014, coinciding with the IMF peak, and the correlation coefficient is 0.63 from 2010 to 2015. The total dipole flux then gradually decreased to the pre-peak level by early 2017 and stayed at the same level after that. The correlation coefficient is 0.74 from 2015 to 2021. The overall time variation of the total dipole flux is well correlated with that of the IMF at 1 AU, and the correlation coefficient is 0.68.

Figure \ref{pic:di_nondi} (B) shows the variation in solar total non-dipole flux in the photosphere as compared to the variation in IMF from 2010 to 2021. The solar total non-dipole flux was derived by calculating Equation (\ref{eq:SHF1}) for $\ell\geq2$.
The total non-dipole flux increased slowly from May 2010 to December 2014 and the correlation coefficient is 0.43 from 2010 to 2015. This trend differs from that of the IMF.
After the peak in December 2014, the total non-dipole flux gradually decreased until April 2020, showing a tendency similar to that of the IMF.
The correlation between the two is $0.42$ for the entire period. However, if we narrow the period to only after 2015, the correlation becomes as high as $0.87$.

\begin{figure}[!]
  \centering
  \includegraphics[width=0.95\textwidth]{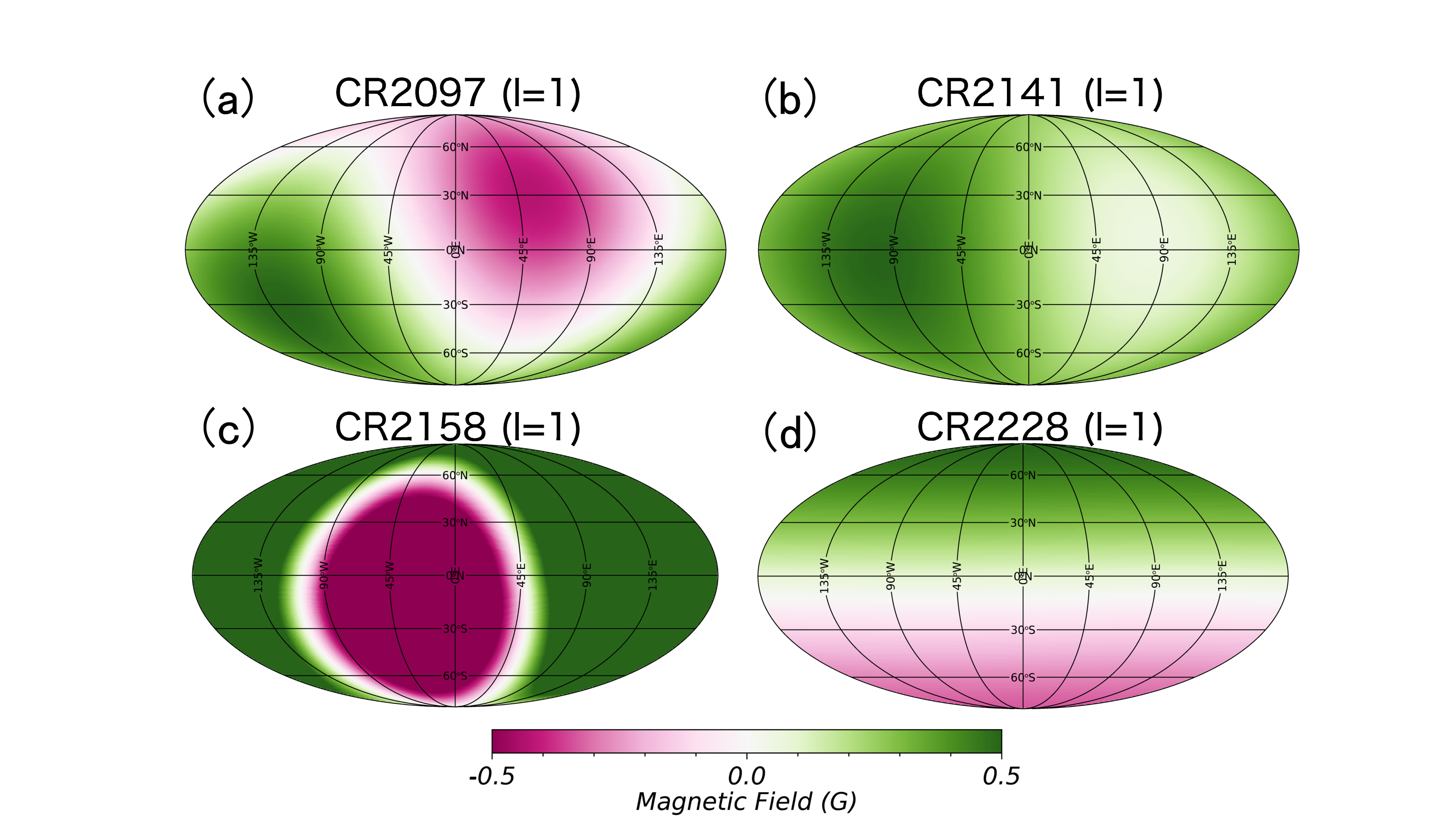}
  \caption{The spatial distribution of the total dipole component on the solar surface. The four maps are derived at the times indicated by the dotted lines in Figure \ref{pic:di_nondi} (A). (a) CR2097 (May 2010 to June 2010), (b) CR2141 (August 2013 to September 2013), (c) CR2158 (December 2014 to January 2015), and (d) CR2228 (in March 2020).}
  \label{pic:2d_l1}
 \end{figure}

\begin{figure}[ht!]
  \centering
  \includegraphics[width=0.95\textwidth,clip]{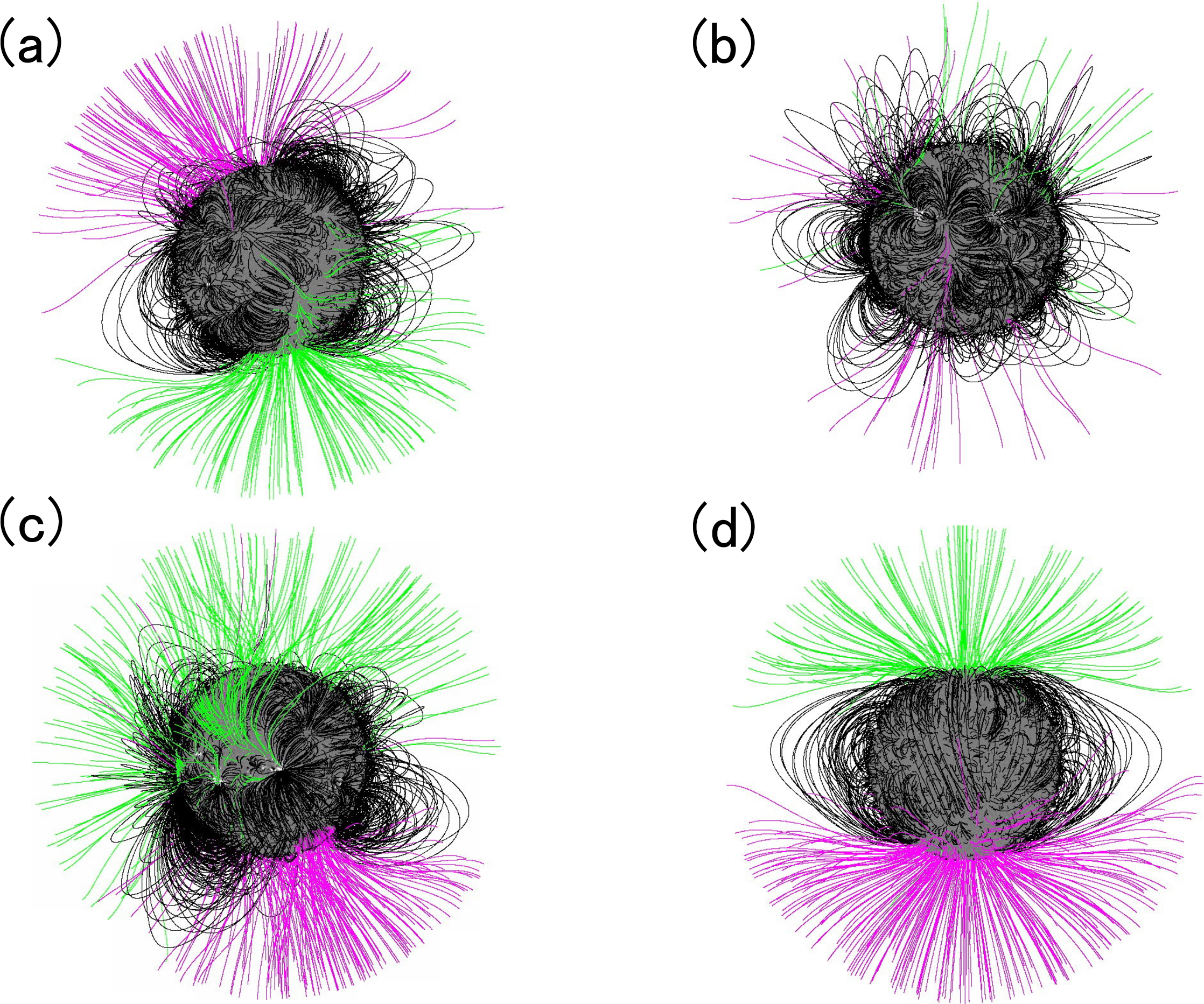}
  \caption{The solar magnetic structures at four times. (a) CR2097 (May 2010 to June 2010), (b) CR2141 (August 2013 to September 2013), (c) CR2158 (December 2014 to January 2015), and (d) CR2228 (in March 2020).}
 \label{pic:fieldline}
 \end{figure}

Figure \ref{pic:2d_l1} shows the total dipole magnetic flux maps (i.e., $\ell=1$) at four timings given in Figure \ref{pic:di_nondi} (A), CR2097 (from May 2010 to June 2010), CR2141 (September 2013), CR2158 (from December 2014 to January 2015), and CR2228 (March 2020), which correspond to the increasing phase, solar maximum, peak time of the IMF (the open flux as well), and minimum, respectively.
The map for CR2097 (upper left) corresponds to the increasing phase between the solar minimum and maximum, and the total dipole component includes both equatorial and axial components. The equatorial dipole is predominant in the map for CR2141 (upper right), i.e., just before the sunspot maximum. The equatorial dipole is still dominant in the map for CR2158 (lower left), just after the maximum. However, the magnetic field strength is ten times larger than that in CR2141. Finally, CR2228 (lower right) exhibits a predominance of the axial dipole at the solar minimum.
Figure \ref{pic:fieldline} shows the structures of the coronal magnetic field using the PFSS model on four occasions.
The dipole structure and its magnitude appear to be different on the four occasions depending on the cycle phase.
Figure \ref{pic:fieldline} (a) shows the spatial distribution of the dipole structure during the ascending phase after the solar minimum. The axis of the dipole is slightly tilted from the rotation axis, indicating the mixture of axial and equatorial dipoles.
At the solar maximum in Figure \ref{pic:fieldline} (b), the axial dipole is unclear.
As shown in Figure \ref{pic:fieldline} (c), when the open flux is at its peak, the equatorial dipole surpasses the axial dipole, and its field strength is stronger than that of (b).
In Figure \ref{pic:fieldline} (d), the total dipole field is stable and the axial dipole is dominant, which is the typical structure in the solar minimum.

\begin{figure}[ht!]
 \centering
 \plotone{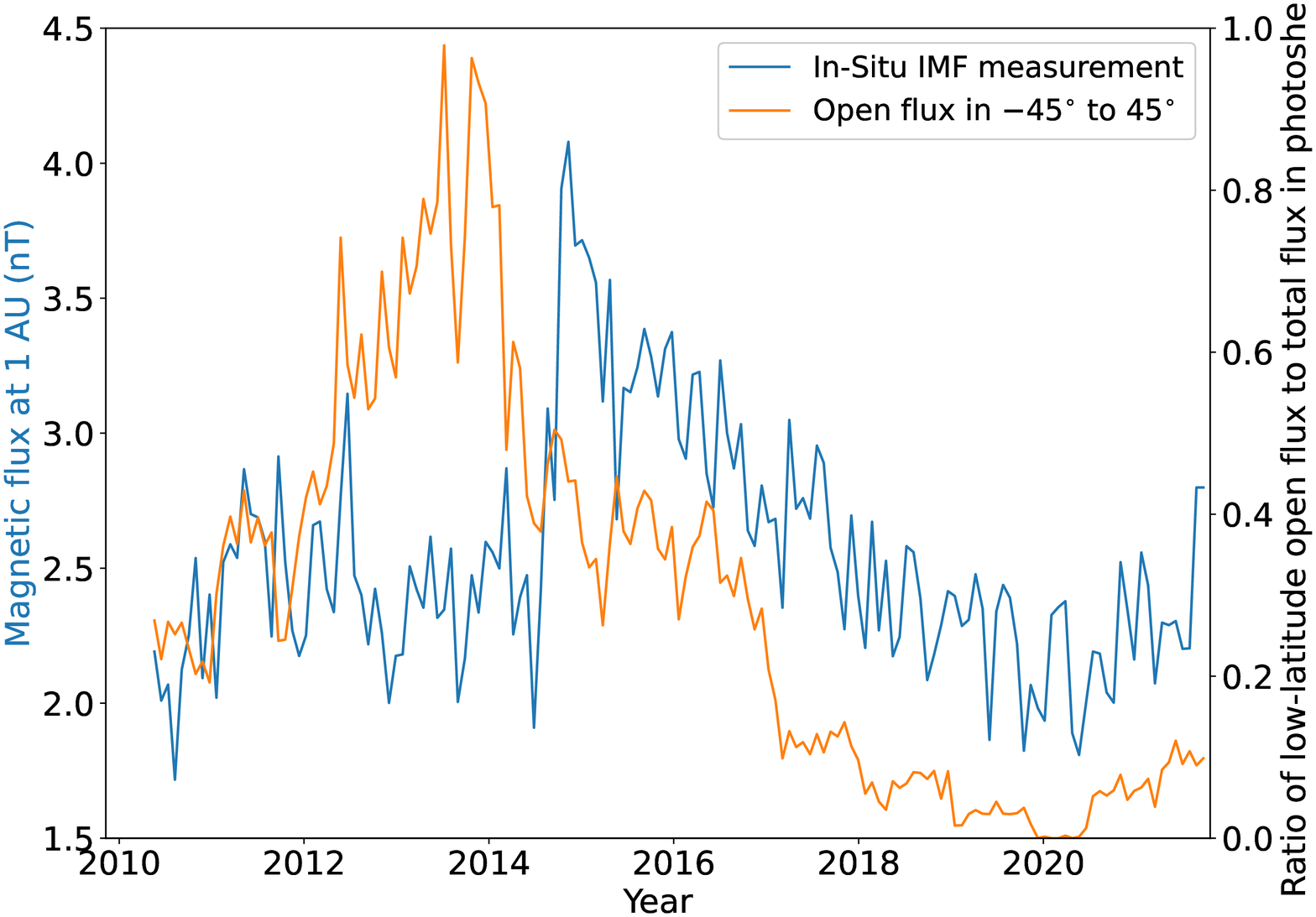}
 \caption{The variations of IMF by in-situ measurements near Earth (blue line, the left axis) and the ratio of the open flux in the low-mid latitudes from $-45^{\circ}$ to $45^{\circ}$ to the total at the photosphere (orange line, the right axis) during 2010 to 2021. Each data point corresponds to the data for the  Carrington rotation.}
 \label{pic:open45-90}
 \end{figure}

The open flux extends mainly from the polar regions during the period when the solar magnetic field displays an axial dipole structure, as shown in Figure \ref{pic:fieldline} (d). Figure \ref{pic:open45-90} shows the temporal evolution of the percentage of the flux originating at low latitudes, defined here as $-45^{\circ}$ to $45^{\circ}$.
The ratio is the smallest in CR2230 (April 2020 to May 2020), and the IMF flux also hits the bottom during the same period. The fraction of low-latitude open flux is less than 10\% in the period after 2017, indicating that most of the open flux originated at higher latitudes.
The photospheric source of the low-latitude open flux increased to 45-50\% from 2014 to 2016 when the magnitude of the IMF and the open flux both increased. In 2012 to 2013, the Sun is at the maximum phase when a significant amount of magnetic flux is distributed at lower latitudes with less flux in the polar regions.
More than 70\% of the open flux originates between $-45^{\circ}$ and $45^{\circ}$.

\begin{figure}[ht!]
 \centering
 \plotone{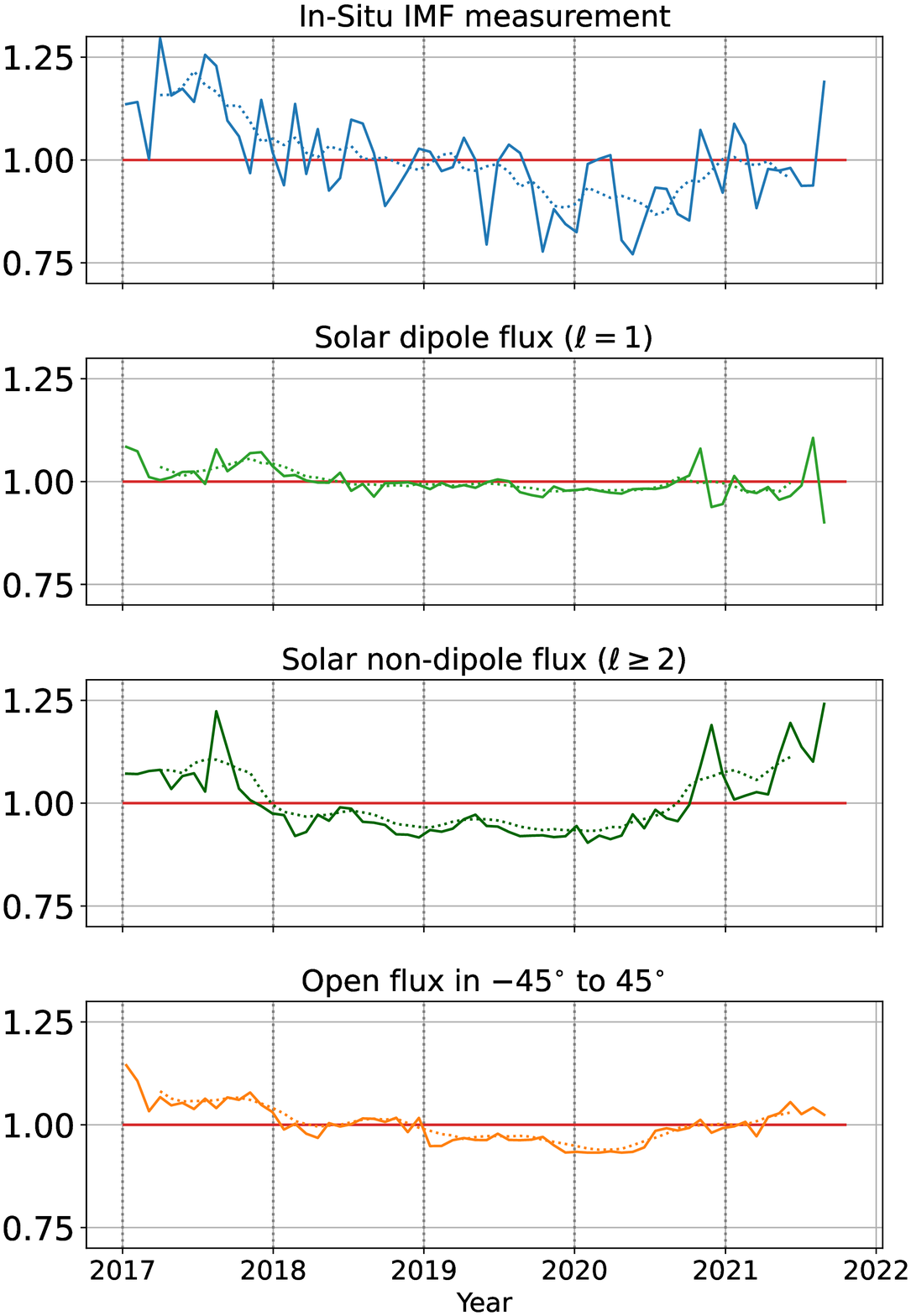}
 \caption{Focus on the period around the solar minimum (January 2017 - October 2021, the horizontal axis) in Figures \ref{pic:di_nondi} (A), \ref{pic:di_nondi} (B), and \ref{pic:open45-90}. The IMF, solar total dipole flux, total non-dipole flux, and open flux in $-45^{\circ}$ to $45^{\circ}$ are normalized over the solar cycle and shifted so that the averages of them during the solar minimum (January 2017 to October 2021) are 1.0 (red solid lines). The dotted lines show their 6 rotation running averages.}
 \label{pic:min}
 \end{figure}

\begin{figure}[ht!]
 \centering
 \plotone{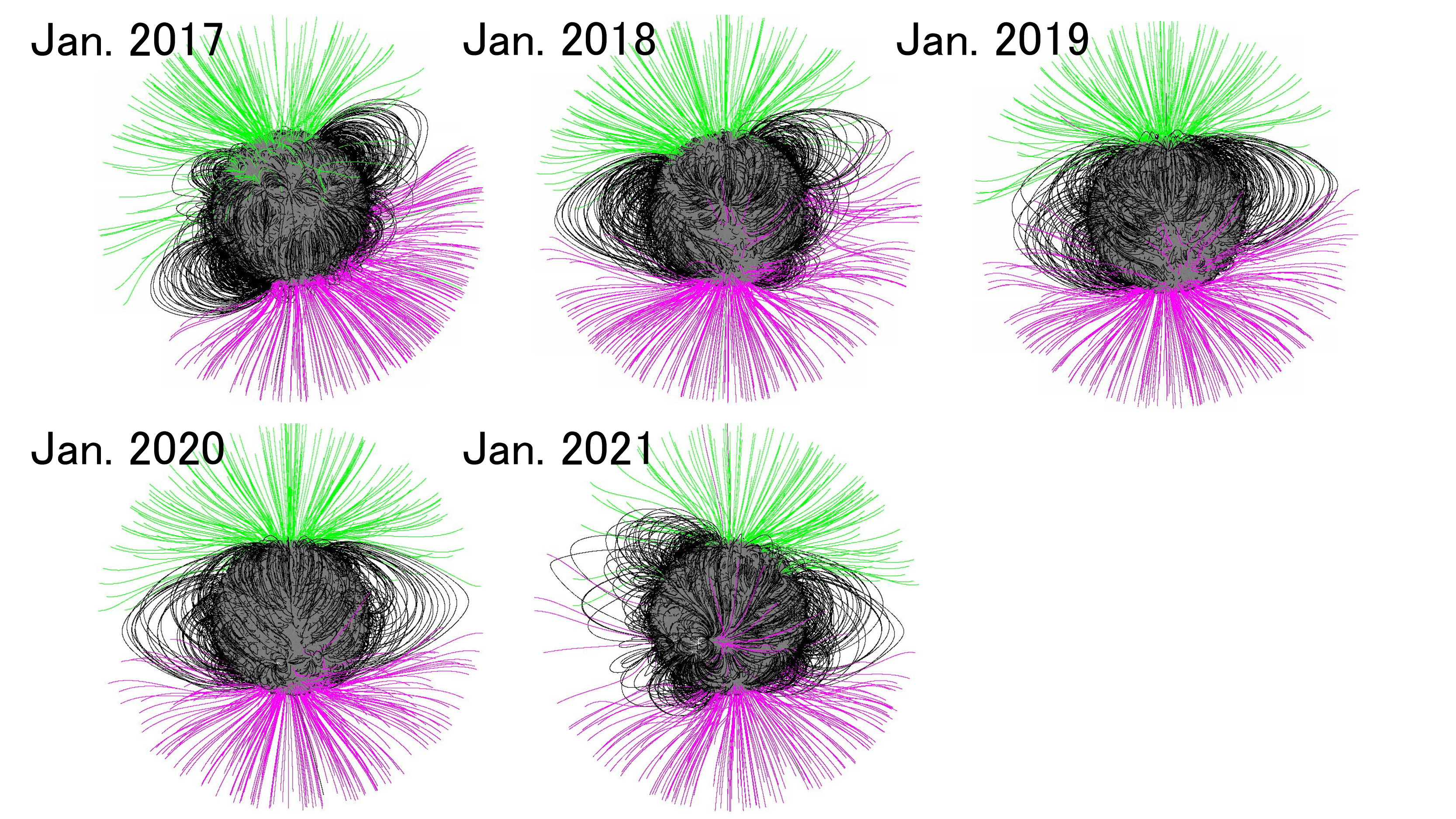}
 \caption{The solar magnetic structures at five times during the solar minimum. January 2017 (CR2186), January 2018 (CR2200), January 2019 (CR2213), January 2020 (CR2226), and January 2021 (CR2239).}
 \label{pic:line_min}
 \end{figure}

Figure \ref{pic:min} shows an enlarged view of the minimum period adopted from Figures \ref{pic:di_nondi} (A), \ref{pic:di_nondi} (B), and \ref{pic:open45-90}.
To compare the magnitude of variations in the period from January 2017 to October 2021, we normalized all the data as follows: First, each data set was normalized by the peak amplitude (maximum minus minimum) in the period from 2010 to 2021. Then, each data was shifted to set the average from January 2017 to October 2021 at 1.0 so that it clearly shows the magnitude of the deviation from the average.
The peak-to-peak value of the IMF variations (blue line) in this period is 35\%; that of the non-dipole (dark green line, $\ell\geq2$) is 18\%; the variation of the low-latitude open flux (orange line) is 15\%; and the variation amplitude of the total dipole flux (green line, $\ell=1$) is only 8\% in Figure \ref{pic:min}.
The standard deviations are 0.12 (IMF), 0.033 (total dipole flux), 0.07 (total non-dipole flux), and 0.05 (the low-latitude open flux). The correlation coefficients with IMF are 0.29 (total dipole flux), 0.44 (total non-dipole flux), and 0.61 (the low-latitude open flux) from 2017 to 2021 (Table \ref{table:pcc}).
The change in the total non-dipole flux was double than that of the total dipole flux. The IMF exhibits change four times larger than that of the total dipole flux. This means that the amplitude of IMF changes in the solar minimum period is closer to that of the total non-dipole flux rather than the total dipole flux.
Figure \ref{pic:line_min} shows the year-by-year evolution of the coronal field structure in January of 2017 - 2021, illustrating that the global structures do not change on a yearly scale during the solar minimum. Most of the open flux is originated from the polar regions. It should be noted that a small amount of open flux is seen from mid-low latitudes except in January 2020 and that the existence of such open flux at low latitudes is reflected in the evolution given by the orange line in Figure \ref{pic:min}.

\section{Discussion}\label{sec:Discussion}
In this study, we derived the temporal evolution of the open flux estimated by the PFSS model over one solar cycle. As a result, we found that the underestimation against the in-situ measurement of the radial IMF near Earth was by a factor of 3 to 4, as previously reported. The IMF suddenly increased in December 2014, which is just after the sunspot maximum, and it decreased gradually over the timescale of a few years. The estimated open flux also showed a similar temporal behavior.
By separating the total dipole components of the coronal magnetic field from the total non-dipole components,
we found that the IMF fluctuates similarly to the non-dipole field in the period before December 2014 (Figure \ref{pic:di_nondi} (B)) and that the IMF behaviors were similar to those of the solar total dipole component when the enhanced IMF was observed from 2014 to 2018 (Figure \ref{pic:di_nondi} (A)).
During that period, the equatorial dipole component was dominant, although the rest of the period showed dominance of the axial component.

This result indicates that the connectivity of the open flux components on the solar surface to the IMF should be examined separately for each phase of the solar cycle. In this section, we discuss the origin of the IMF on the solar surface in relation to the cycle phases.

\subsection{Open flux during solar maximum}\label{sec:dis_max}
\begin{figure}[ht!]
  \centering
  \includegraphics[width=0.95\columnwidth,clip]{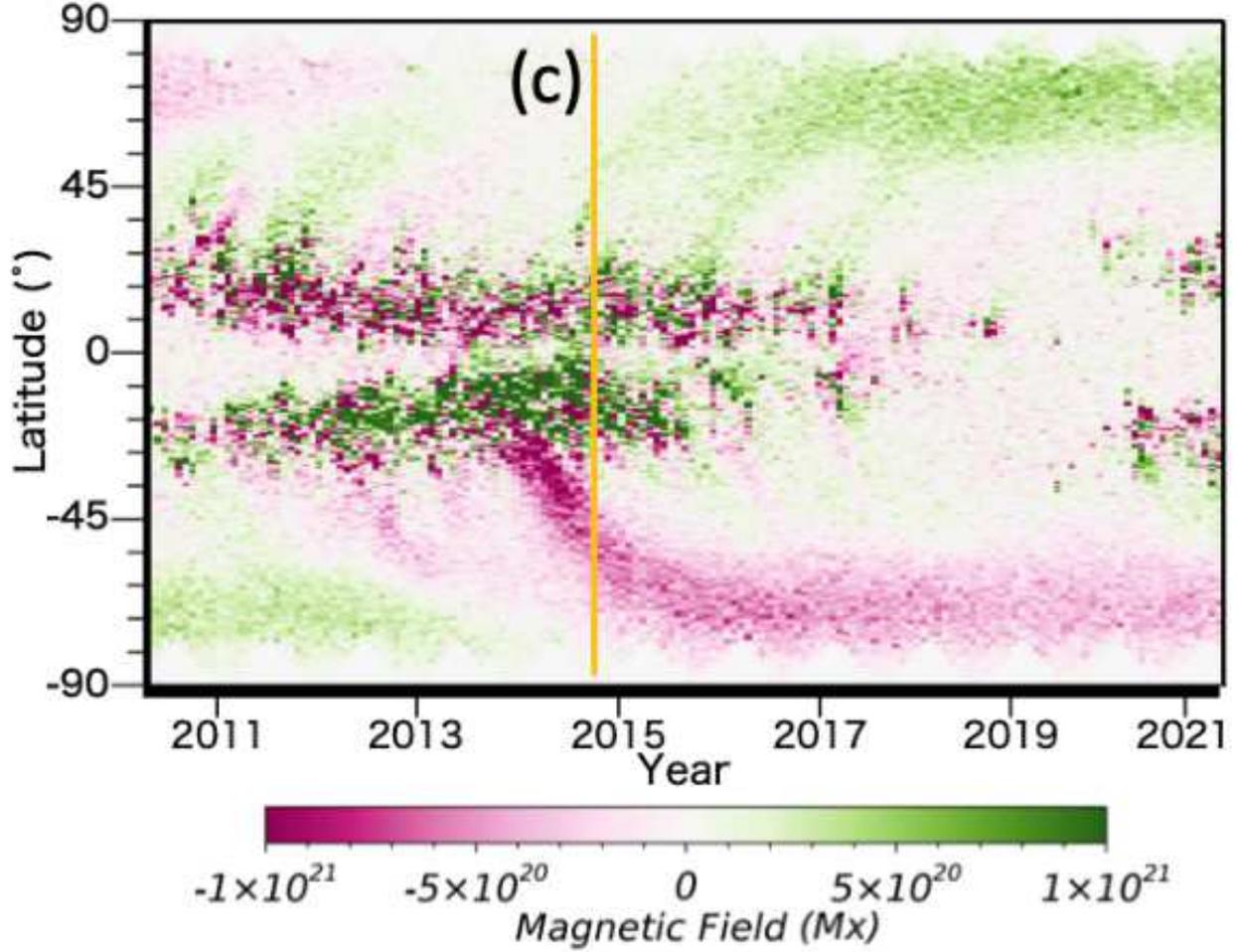}
  \caption{The butterfly diagram from May 2010 to October 2021. The magnetic field in the figure is represented by magnetic flux in a 1-deg grid. The amount of open flux reaches its peak when the magnetic field in the active region is diffused to around $\pm45^{\circ}$ latitudes. From 2014 to 2015, the magnetic field in the southern hemisphere is predominant (yellow line), as shown correspondingly in Figure \ref{pic:2d_l1} (c).}
  \label{pic:2dfly}
 \end{figure}

The comparison between Figures \ref{pic:di_nondi} (A) and \ref{pic:di_nondi} (B) suggests that the IMF variation is associated with the solar total dipole flux, rather than the total non-dipole flux in the solar maximum period (2010 to 2016).
The dipole structure is well observed during the minimum phase, as shown in Figure \ref{pic:fieldline} (d), because of the intense unipolar flux in the polar regions. But magnetic flux at the polar regions, which was weak during the solar maximum from 2012 to 2014, as shown in the butterfly diagram in Figure \ref{pic:2dfly}. Therefore, the total dipole component should be oriented away from the rotational axis in the maximum phase.
The dipole field can be divided into the equatorial dipole, $(\ell, m)=(1, \pm1$), and the axial dipole, $(\ell, m)=(1, 0)$.
When the axial dipole is dominant, the Sun has a dipole field structure with open flux concentrated at the poles as shown in Figure \ref{pic:fieldline} (d).
This structure was not formed when the equatorial dipole takes over.
In the solar maximum, the equatorial dipole was predominant, as shown in Figure \ref{pic:2d_l1} (b) and (c).
Therefore, in the maximum, even though the global magnetic field of the Sun does not have a dipole structure that is oriented in the rotational axis, the dipole field component still dominates in the form of the equatorial dipole.

The IMF radial flux increased in December 2014, which is after the solar maximum and at the beginning of the declining phase in terms of the sunspot number, and this increase in IMF coincides with the increase in the equatorial dipole flux. Then, what causes the dominance of the equatorial dipole field?
Here we point out the possibility that this is related to the diffusion of active regions, namely, the equatorial dipole and IMF peaks coincide with the period when active regions diffuse toward the poles (Figure \ref{pic:2dfly}), which has been attributed to the meridional circulation (\citealp{2016STP.....2a...3M}; \citealp{2017SoPh..292..175G}). The diffusion of active regions also occurs in the longitudinal direction owing to the supergranulation and differential rotation, which have an inverse cascade effect, producing the equatorial dipole $(\ell, m)=(1, \pm1)$ from the active regions $(\ell\geq2)$. As the emergence and latitudinal transport of active regions progress, the polar flux reversal eventually occurs (Babcock-Leighton mechanism: \citealp{1961ApJ...133..572B}; \citealp{1969ApJ...156....1L}) and the new axial dipole structure $(\ell, m)=(1, 0)$ is created, as suggested by \citet{2000GeoRL..27..505W}. The dipole structure stays at a constant level after 2017, as seen in Figures \ref{pic:di_nondi} (A) and \ref{pic:fieldline}. At this time, the unipolar flux region in the pole was formed in both hemispheres after the active region remnants moved toward the poles (Figure \ref{pic:2dfly}).
We consider the scenario described above as a possible cause of the dominance of the equatorial dipole field. However, the exact reason for the sudden increase of the equatorial dipole remains unclear. According to \citet{2014SSRv..186..387W}, depending on the emergence locations of sunspot groups, the equatorial dipole can be intensified, such as when two bipolar sunspot groups are located $180^{\circ}$ apart and on opposite sides of the equator. This possibility has to be explored by the numerical modeling in the future.
Also, it is recently shown by \citet{2023arXiv230407649A} that the amount of open flux depends on the direction in which the magnetic field lines are traced, such as inward (from the outer boundary to the photosphere), outward (from the photosphere to the outer boundary), and bi-directional. Since only the inward tracing was adopted in this study, the next step would be to use different tracing methods and quantitatively investigate their dependence.

\subsection{Open flux during solar minimum}\label{sec:dis_min}
Figure \ref{pic:min} shows that the behavior of the long-term changes in the total non-dipole flux ($\ell\geq2$) is more similar to that of the IMF than that of the total dipole flux ($\ell=1$) during the solar minimum. The magnitudes of the changes during 2017 to 2020 were  0.77 to 1.30, i.e., 35\% for the IMF,  0.90 to 1.11, i.e., 8\% for the total  dipole flux, and 0.90 to 1.24, i.e., 18\% for the total non-dipole flux.
The dipole field structure is globally stable for several years during the solar minimum, but the open flux extending from mid-low latitudes is relatively variable as shown in Figure \ref{pic:line_min}. Thus, we conclude that the evolution of IMF in the minimum was caused by variations in the total non-dipole flux rather than in the total dipole flux.

We also discuss the influence of the photospheric magnetic field in the polar regions, which has been considered to be one of the primary causes of the open flux problem. Our result suggests that the underestimation of the polar magnetic field does not fully resolve the open flux problem during 2017 to 2022.
During the minimum period (2017 to 2022), the open flux is rooted in the polar regions, and the photospheric magnetic field was most concentrated at the high latitude regions in December 2019 (Figure \ref{pic:open45-90}). Therefore, the estimated open flux (Figure \ref{pic:problem} red line) should be the worst estimate in December 2019 if the polar magnetic field was underestimated and caused the open flux problem.
However, as shown in Figure \ref{pic:problem}, within the period from 2017 to 2022, the open flux value was closest to the IMF from December 2019 to January 2020. Therefore, it is suggested that the open flux problem has other causes than the underestimation of the polar magnetic field.

\subsection{The strength of the open flux and IMF}
Our approach of calculating the IMF strength is a simple method, in which we average the IMF data of 1-hour resolution over each solar rotation. According to \citet{2017JGRA..12210980O}, this method may overestimate the IMF by more than 20-50\% in comparison to the most reliable method using the electron strahl. Also, our method may double count the switchbacks of the IMF that occurs near the satellite, and \citet{2022SoPh..297...82F}, show that averaging of 20-hour data is comparable with the method using electron strahl. Although we plan to use better IMF data in the future studies, neither of these methods are sufficient to solve the open flux problem.

Figures \ref{pic:open} (A) and \ref{pic:open} (B) suggest that the temporal variations of open flux track IMF variations, but the absolute value was underestimated during the solar maximum.
This indicates that the strong magnetic field at the low latitudes (e.g., sunspot magnetic field) may be underestimated rather than the polar field. The magnetic field that creates the equatorial dipole, $(\ell, m)=(1, \pm1)$, which dominates during the solar maximum, may be underestimated and cause the open flux problem.
In the solar minimum, contrary to the solar maximum, the axial dipole is dominant. However, it is likely that the variation of the IMF is mainly produced by the total non-dipole component: The total non-dipole flux ($10^{23}$ Mx) is an order of magnitude larger than the total dipole flux ($10^{22}$ Mx) as shown in Figures \ref{pic:di_nondi} (A) and \ref{pic:di_nondi} (B). The equatorial dipole that has developed during the solar maximum was reduced and stable in the solar minimum, while the axial dipole was stable with the photospheric polar magnetic field, which has surpassed the equatorial dipole. We have shown that the total non-dipole component (i.e., the higher-order component) can produce the open field and contribute to the variation in the IMF. Only a limited amount of the total non-dipole component may be open to the interplanetary space, as inferred from Figure \ref{pic:line_min}. However, the total non-dipole component has at least one order of magnitude larger flux than the total dipole component does and may contribute to the underestimation of the open flux. Therefore, we should consider the possibility that the higher-order components of the photospheric magnetic field may be critical in solving the open flux problem.

\section{Conclusion}\label{sec:Conclusion}
The open flux problem is now considered one of the major remaining issues in solar and heliophysics because it highlights the fact that we lack knowledge of the connection between the solar magnetic field and IMF. In this study, we extrapolated the coronal magnetic field from the solar photospheric magnetic field using the PFSS model and decomposed it into the dipole ($\ell=1$) and non-dipole ($\ell\geq2$) components. We investigated the relationship between the IMF and the solar magnetic field by comparing the time variability of each component with that of the IMF.
We found that the IMF rapidly increased to its peak at CR2157 (November 2014 to December 2014) seven months after the sunspot number maximum (April 2014). This period coincides with when the solar total dipole flux reaches its maximum. The equatorial dipole flux was dominant and rapidly increased, which corresponds to the moving of active region remnants toward the south pole from April 2014 to December 2014.
On the other hand, in the minimum phase, the IMF decreased as well as the sunspot number did, reaching the minimum in December 2019. After December 2019, the IMF slowly increased. From 2017 to 2021, the variation amplitude of IMF was about 35\%. In the same period, the total non-dipole flux variation was 18\%, the variations in low-latitude open flux was 15\%, while the total dipole flux variation was 8\%.

We discussed the relationship between the IMF and the open flux variation, and the conclusions are summarized as follows.
In the solar maximum, active regions diffuse their magnetic field toward the poles owing to meridional circulation and in the longitudinal direction owing to differential rotation. The solar equatorial dipole flux, $(\ell, m)=(1, \pm1)$, becomes dominant in this process. The equatorial dipole flux may reach the source surface and contribute to the open flux, which extends further into the IMF near Earth. This is evident by the fact that the IMF increased rapidly in December 2014 in synchronization with the development of an equatorial dipole component.
The axial dipole $(\ell, m)=(1, 0)$ in the solar minimum is stable and forms a global dipole field structure. On the other hand, the IMF variation is large in 2017-2021 and matches the variation in the total non-dipole flux ($\ell\geq2$).

The open flux inferred from the solar surface is underestimated by a factor of 3 to 4 compared to the IMF measured near Earth.
Our results show that the IMF peak is synchronous with the sudden enhancement of equatorial dipole flux. We also found that the magnitude of the total non-dipole flux is at least one order larger than that of the total dipole flux, which indicates that the total non-dipole flux may contribute significantly to the open flux. Therefore, we suggest that a detailed study of the equatorial dipole flux and total non-dipole flux is a key to approaching the open flux problem in the future.

\section*{Acknowledgements}
The data courtesy of NASA/SDO and the HMI science team. We thank Joint Science Operations Center at Stanford for providing HMI synoptic maps.
We acknowledge use of NASA/GSFC's Space Physics Data Facility's OMNIWeb service and OMNI data.
We are grateful to H. Iijima of Nagoya University for the useful comments.
We thank the anonymous referee for the careful and valuable comments.
This work was supported by JSPS KAKENHI Grant Nos. JP20KK0072 (PI: S. Toriumi), JP21H01124 (PI: T. Yokoyama), JP21H04492 (PI: K. Kusano), and JP18H05234 (PI: Y. Katsukawa).

\appendix
\section{Spherical harmonic function}\label{appendix}
The expression for expanding $Y$ to the spherical harmonic can be expressed as follows:
\begin{equation}
  Y^{m}_{\ell}(\theta, \phi)=C^{m}_{\ell}P^{m}_{\ell}(\cos\theta)e^{im\phi},  \label{eq:SHF1}
\end{equation}
where $\ell$ and $m$ are integers with $0\leq \ell$ and $-\ell \leq m < \ell$, equation $P^{m}_{\ell}$ is the associated Legendre functions, and constant $C^{m}_{\ell}$ is defined as
\begin{equation}
  C^{m}_{\ell}=(-1)^{m} \biggl[\frac{2\ell+1}{4\pi} \frac{(\ell-m)!}{(\ell+m)!}\biggr]^{\frac{1}{2}}.
  \label{eq:SHF3}
\end{equation}
Expanding the function $f(\theta, \phi)$ over $Y^{m}_{\ell}$, we obtained the coefficients $A^{m}_{\ell}$ such that
\begin{equation}
  f(\theta, \phi)=\sum_{\ell=0}^\infty \sum_{m=-l}^\infty A_{\ell}^{m}Y_{l}^{m} (\theta, \phi),
  \label{eq:SHF2}
\end{equation}
where $\theta$ and $\phi$ are angles in the spherical polar coordinate system.



\bibliography{sample631}{}

\begin{thebibliography}{}
\expandafter\ifx\csname natexlab\endcsname\relax\def\natexlab#1{#1}\fi
\providecommand{\url}[1]{\href{#1}{#1}}
\providecommand{\dodoi}[1]{doi:~\href{http://doi.org/#1}{\nolinkurl{#1}}}
\providecommand{\doeprint}[1]{\href{http://ascl.net/#1}{\nolinkurl{http://ascl.net/#1}}}
\providecommand{\doarXiv}[1]{\href{https://arxiv.org/abs/#1}{\nolinkurl{https://arxiv.org/abs/#1}}}

\bibitem[{{Altschuler} \& {Newkirk}(1969)}]{1969SoPh....9..131A}
{Altschuler}, M.~D., \& {Newkirk}, G. 1969, \solphys, 9, 131,
  \dodoi{10.1007/BF00145734}

\bibitem[{{Arge} {et~al.}(2023){Arge}, {Leisner}, {Wallace}, \&
  {Henney}}]{2023arXiv230407649A}
{Arge}, C.~N., {Leisner}, A., {Wallace}, S., \& {Henney}, C.~J. 2023, arXiv
  e-prints, arXiv:2304.07649, \dodoi{10.48550/arXiv.2304.07649}

\bibitem[{{Asvestari} {et~al.}(2019){Asvestari}, {Heinemann}, {Temmer},
  {Pomoell}, {Kilpua}, {Magdalenic}, \& {Poedts}}]{2019JGRA..124.8280A}
{Asvestari}, E., {Heinemann}, S.~G., {Temmer}, M., {et~al.} 2019, Journal of
  Geophysical Research (Space Physics), 124, 8280, \dodoi{10.1029/2019JA027173}

\bibitem[{{Babcock}(1961)}]{1961ApJ...133..572B}
{Babcock}, H.~W. 1961, \apj, 133, 572, \dodoi{10.1086/147060}

\bibitem[{{Badman} {et~al.}(2021){Badman}, {Bale}, {Rouillard}, {Bowen},
  {Bonnell}, {Goetz}, {Harvey}, {MacDowall}, {Malaspina}, \&
  {Pulupa}}]{2021A&A...650A..18B}
{Badman}, S.~T., {Bale}, S.~D., {Rouillard}, A.~P., {et~al.} 2021, \aap, 650,
  A18, \dodoi{10.1051/0004-6361/202039407}

\bibitem[{{Bernath} {et~al.}(2005){Bernath}, {McElroy}, {Abrams}, {Boone},
  {Butler}, {Camy-Peyret}, {Carleer}, {Clerbaux}, {Coheur}, {Colin}, {DeCola},
  {DeMazi{\`e}re}, {Drummond}, {Dufour}, {Evans}, {Fast}, {Fussen}, {Gilbert},
  {Jennings}, {Llewellyn}, {Lowe}, {Mahieu}, {McConnell}, {McHugh}, {McLeod},
  {Michaud}, {Midwinter}, {Nassar}, {Nichitiu}, {Nowlan}, {Rinsland}, {Rochon},
  {Rowlands}, {Semeniuk}, {Simon}, {Skelton}, {Sloan}, {Soucy}, {Strong},
  {Tremblay}, {Turnbull}, {Walker}, {Walkty}, {Wardle}, {Wehrle}, {Zander}, \&
  {Zou}}]{2005GeoRL..3215S01B}
{Bernath}, P.~F., {McElroy}, C.~T., {Abrams}, M.~C., {et~al.} 2005, \grl, 32,
  L15S01, \dodoi{10.1029/2005GL022386}

\bibitem[{{Decraemer} {et~al.}(2019){Decraemer}, {Zhukov}, \& {Van
  Doorsselaere}}]{2019ApJ...883..152D}
{Decraemer}, B., {Zhukov}, A.~N., \& {Van Doorsselaere}, T. 2019, \apj, 883,
  152, \dodoi{10.3847/1538-4357/ab3b58}

\bibitem[{{Frost} {et~al.}(2022){Frost}, {Owens}, {Macneil}, \&
  {Lockwood}}]{2022SoPh..297...82F}
{Frost}, A.~M., {Owens}, M., {Macneil}, A., \& {Lockwood}, M. 2022, \solphys,
  297, 82, \dodoi{10.1007/s11207-022-02004-6}

\bibitem[{{Golubeva} \& {Mordvinov}(2017)}]{2017SoPh..292..175G}
{Golubeva}, E.~M., \& {Mordvinov}, A.~V. 2017, \solphys, 292, 175,
  \dodoi{10.1007/s11207-017-1200-6}

\bibitem[{{Heinemann} {et~al.}(2019){Heinemann}, {Temmer}, {Heinemann},
  {Dissauer}, {Samara}, {Jer{\v{c}}i{\'c}}, {Hofmeister}, \&
  {Veronig}}]{2019SoPh..294..144H}
{Heinemann}, S.~G., {Temmer}, M., {Heinemann}, N., {et~al.} 2019, \solphys,
  294, 144, \dodoi{10.1007/s11207-019-1539-y}

\bibitem[{{Ito} {et~al.}(2010){Ito}, {Tsuneta}, {Shiota}, {Tokumaru}, \&
  {Fujiki}}]{2010ApJ...719..131I}
{Ito}, H., {Tsuneta}, S., {Shiota}, D., {Tokumaru}, M., \& {Fujiki}, K. 2010,
  \apj, 719, 131, \dodoi{10.1088/0004-637X/719/1/131}

\bibitem[{King \& Papitashvili(2005)}]{https://doi.org/10.1029/2004JA010649}
King, J.~H., \& Papitashvili, N.~E. 2005, Journal of Geophysical Research:
  Space Physics, 110, \dodoi{https://doi.org/10.1029/2004JA010649}

\bibitem[{{Koutchmy} \& {Livshits}(1992)}]{1992SSRv...61..393K}
{Koutchmy}, S., \& {Livshits}, M. 1992, \ssr, 61, 393,
  \dodoi{10.1007/BF00222313}

\bibitem[{{Krieger} {et~al.}(1973){Krieger}, {Timothy}, \&
  {Roelof}}]{1973SoPh...29..505K}
{Krieger}, A.~S., {Timothy}, A.~F., \& {Roelof}, E.~C. 1973, \solphys, 29, 505,
  \dodoi{10.1007/BF00150828}

\bibitem[{{Leighton}(1969)}]{1969ApJ...156....1L}
{Leighton}, R.~B. 1969, \apj, 156, 1, \dodoi{10.1086/149943}

\bibitem[{{Lepping} {et~al.}(1995){Lepping}, {Ac{\~{u}}na}, {Burlaga},
  {Farrell}, {Slavin}, {Schatten}, {Mariani}, {Ness}, {Neubauer}, {Whang},
  {Byrnes}, {Kennon}, {Panetta}, {Scheifele}, \&
  {Worley}}]{1995SSRv...71..207L}
{Lepping}, R.~P., {Ac{\~{u}}na}, M.~H., {Burlaga}, L.~F., {et~al.} 1995, \ssr,
  71, 207, \dodoi{10.1007/BF00751330}

\bibitem[{{Linker} {et~al.}(2017){Linker}, {Caplan}, {Downs}, {Riley}, {Mikic},
  {Lionello}, {Henney}, {Arge}, {Liu}, {Derosa}, {Yeates}, \&
  {Owens}}]{2017ApJ...848...70L}
{Linker}, J.~A., {Caplan}, R.~M., {Downs}, C., {et~al.} 2017, \apj, 848, 70,
  \dodoi{10.3847/1538-4357/aa8a70}

\bibitem[{{Linker} {et~al.}(2021){Linker}, {Heinemann}, {Temmer}, {Owens},
  {Caplan}, {Arge}, {Asvestari}, {Delouille}, {Downs}, {Hofmeister}, {Jebaraj},
  {Madjarska}, {Pinto}, {Pomoell}, {Samara}, {Scolini}, \&
  {Vr{\v{s}}nak}}]{2021ApJ...918...21L}
{Linker}, J.~A., {Heinemann}, S.~G., {Temmer}, M., {et~al.} 2021, \apj, 918,
  21, \dodoi{10.3847/1538-4357/ac090a}

\bibitem[{{Mackay} \& {Yeates}(2012)}]{2012LRSP....9....6M}
{Mackay}, D.~H., \& {Yeates}, A.~R. 2012, Living Reviews in Solar Physics, 9,
  6, \dodoi{10.12942/lrsp-2012-6}

\bibitem[{{McComas} {et~al.}(1998){McComas}, {Bame}, {Barraclough}, {Feldman},
  {Funsten}, {Gosling}, {Riley}, {Skoug}, {Balogh}, {Forsyth}, {Goldstein}, \&
  {Neugebauer}}]{1998GeoRL..25....1M}
{McComas}, D.~J., {Bame}, S.~J., {Barraclough}, B.~L., {et~al.} 1998, \grl, 25,
  1, \dodoi{10.1029/97GL03444}

\bibitem[{{Miki{\'c}} {et~al.}(2018){Miki{\'c}}, {}, {Downs}, {Linker},
  {Caplan}, {Mackay}, {Upton}, {Riley}, {Lionello}, {T{\"o}r{\"o}k}, {Titov},
  {Wijaya}, {Druckm{\"u}ller}, {Pasachoff}, \& {Carlos}}]{2018NatAs...2..913M}
{Miki{\'c}}, {}, Z., {Downs}, C., {et~al.} 2018, Nature Astronomy, 2, 913,
  \dodoi{10.1038/s41550-018-0562-5}

\bibitem[{{Mordvinov} {et~al.}(2016){Mordvinov}, {Pevtsov}, {Bertello}, \&
  {Petri}}]{2016STP.....2a...3M}
{Mordvinov}, A., {Pevtsov}, A., {Bertello}, L., \& {Petri}, G. 2016,
  Solar-Terrestrial Physics, 2, 3, \dodoi{10.12737/16356}

\bibitem[{{Owens} {et~al.}(2008){Owens}, {Arge}, {Crooker}, {Schwadron}, \&
  {Horbury}}]{2008JGRA..11312103O}
{Owens}, M.~J., {Arge}, C.~N., {Crooker}, N.~U., {Schwadron}, N.~A., \&
  {Horbury}, T.~S. 2008, Journal of Geophysical Research (Space Physics), 113,
  A12103, \dodoi{10.1029/2008JA013677}

\bibitem[{{Owens} \& {Forsyth}(2013)}]{2013LRSP...10....5O}
{Owens}, M.~J., \& {Forsyth}, R.~J. 2013, Living Reviews in Solar Physics, 10,
  5, \dodoi{10.12942/lrsp-2013-5}

\bibitem[{{Owens} {et~al.}(2017){Owens}, {Lockwood}, {Riley}, \&
  {Linker}}]{2017JGRA..12210980O}
{Owens}, M.~J., {Lockwood}, M., {Riley}, P., \& {Linker}, J. 2017, Journal of
  Geophysical Research (Space Physics), 122, 10,980,
  \dodoi{10.1002/2017JA024631}

\bibitem[{{Parker}(1958)}]{1958ApJ...128..664P}
{Parker}, E.~N. 1958, \apj, 128, 664, \dodoi{10.1086/146579}

\bibitem[{{Pasachoff} {et~al.}(2015){Pasachoff}, {Ru{\v{s}}in}, {Saniga},
  {Babcock}, {Lu}, {Davis}, {Dantowitz}, {Gaintatzis}, {Seiradakis},
  {Voulgaris}, {Seaton}, \& {Shiota}}]{2015ApJ...800...90P}
{Pasachoff}, J.~M., {Ru{\v{s}}in}, V., {Saniga}, M., {et~al.} 2015, \apj, 800,
  90, \dodoi{10.1088/0004-637X/800/2/90}

\bibitem[{{Pesnell} {et~al.}(2012){Pesnell}, {Thompson}, \&
  {Chamberlin}}]{2012SoPh..275....3P}
{Pesnell}, W.~D., {Thompson}, B.~J., \& {Chamberlin}, P.~C. 2012, \solphys,
  275, 3, \dodoi{10.1007/s11207-011-9841-3}

\bibitem[{{Riley} {et~al.}(2019){Riley}, {Linker}, {Mikic}, {Caplan}, {Downs},
  \& {Thumm}}]{2019ApJ...884...18R}
{Riley}, P., {Linker}, J.~A., {Mikic}, Z., {et~al.} 2019, \apj, 884, 18,
  \dodoi{10.3847/1538-4357/ab3a98}

\bibitem[{{Riley} {et~al.}(2021){Riley}, {Lionello}, {Caplan}, {Downs},
  {Linker}, {Badman}, \& {Stevens}}]{2021A&A...650A..19R}
{Riley}, P., {Lionello}, R., {Caplan}, R.~M., {et~al.} 2021, \aap, 650, A19,
  \dodoi{10.1051/0004-6361/202039815}

\bibitem[{{Schatten} {et~al.}(1969){Schatten}, {Wilcox}, \&
  {Ness}}]{1969SoPh....6..442S}
{Schatten}, K.~H., {Wilcox}, J.~M., \& {Ness}, N.~F. 1969, \solphys, 6, 442,
  \dodoi{10.1007/BF00146478}

\bibitem[{{Scherrer} {et~al.}(2012){Scherrer}, {Schou}, {Bush}, {Kosovichev},
  {Bogart}, {Hoeksema}, {Liu}, {Duvall}, {Zhao}, {Title}, {Schrijver},
  {Tarbell}, \& {Tomczyk}}]{2012SoPh..275..207S}
{Scherrer}, P.~H., {Schou}, J., {Bush}, R.~I., {et~al.} 2012, \solphys, 275,
  207, \dodoi{10.1007/s11207-011-9834-2}

\bibitem[{{Tsuneta} {et~al.}(2008){Tsuneta}, {Ichimoto}, {Katsukawa}, {Lites},
  {Matsuzaki}, {Nagata}, {Orozco Su{\'a}rez}, {Shimizu}, {Shimojo}, {Shine},
  {Suematsu}, {Suzuki}, {Tarbell}, \& {Title}}]{2008ApJ...688.1374T}
{Tsuneta}, S., {Ichimoto}, K., {Katsukawa}, Y., {et~al.} 2008, \apj, 688, 1374,
  \dodoi{10.1086/592226}

\bibitem[{{Wallace} {et~al.}(2019){Wallace}, {Arge}, {Pattichis},
  {Hock-Mysliwiec}, \& {Henney}}]{2019SoPh..294...19W}
{Wallace}, S., {Arge}, C.~N., {Pattichis}, M., {Hock-Mysliwiec}, R.~A., \&
  {Henney}, C.~J. 2019, \solphys, 294, 19, \dodoi{10.1007/s11207-019-1402-1}

\bibitem[{{Wang}(2014)}]{2014SSRv..186..387W}
{Wang}, Y.~M. 2014, \ssr, 186, 387, \dodoi{10.1007/s11214-014-0051-9}

\bibitem[{{Wang} {et~al.}(2000){Wang}, {Lean}, \&
  {Sheeley}}]{2000GeoRL..27..505W}
{Wang}, Y.~M., {Lean}, J., \& {Sheeley}, N.~R., J. 2000, \grl, 27, 505,
  \dodoi{10.1029/1999GL010744}

\bibitem[{{Wang} \& {Sheeley}(1995)}]{1995ApJ...447L.143W}
{Wang}, Y.~M., \& {Sheeley}, N.~R., J. 1995, \apjl, 447, L143,
  \dodoi{10.1086/309578}

\bibitem[{{Wang} {et~al.}(2022){Wang}, {Ulrich}, \&
  {Harvey}}]{2022ApJ...926..113W}
{Wang}, Y.~M., {Ulrich}, R.~K., \& {Harvey}, J.~W. 2022, \apj, 926, 113,
  \dodoi{10.3847/1538-4357/ac4491}

\bibitem[{{Webb} \& {Howard}(2012)}]{2012LRSP....9....3W}
{Webb}, D.~F., \& {Howard}, T.~A. 2012, Living Reviews in Solar Physics, 9, 3,
  \dodoi{10.12942/lrsp-2012-3}

\bibitem[{{Yeates}(2014)}]{2014SoPh..289..631Y}
{Yeates}, A.~R. 2014, \solphys, 289, 631, \dodoi{10.1007/s11207-013-0301-0}

\end{thebibliography}
\bibliographystyle{aasjournal}



\end{document}